# Polydispersity analysis of Taylor dispersion data: the cumulant method

*Luca Cipelletti*[1], *Jean-Philippe Biron*[2], *Michel Martin*[3], *Hervé Cottet*[2]

[1]Laboratoire Charles Coulomb (L2C, UMR 5221 CNRS-Université Montpellier 2), Place Eugène Bataillon, F-34095 Montpellier Cedex 5, France

[2]Institut des Biomolécules Max Mousseron (IBMM, UMR 5247 CNRS-Université de Montpellier 1- Université de Montpellier 2), Place Eugène Bataillon, CC 1706, F-34095 Montpellier Cedex 5, France

[3]Ecole Supérieure de Physique et de Chimie Industrielles, Laboratoire de Physique et Mécanique des Milieux Hétérogènes (PMMH, UMR 7636 CNRS, ESPCI-ParisTech, Université Pierre et Marie Curie, Université Paris-Diderot), 10 rue Vauquelin, F-75231 Paris Cedex 05, France

TITLE RUNNING HEAD. Polydispersity analysis of Taylor dispersion data.

[*]CORRESPONDING AUTHORS

Tel: +33 4 6714 3589, Fax : +33 4 67 14 34 98 E-mail: Luca.Cipelletti@univ-montp2.fr

Tel: +33 4 6714 3427, Fax: +33 4 6763 1046. E-mail: hcottet@univ-montp2.fr

**ABSTRACT**

Taylor dispersion analysis is an increasingly popular characterization method that measures the diffusion coefficient, and hence the hydrodynamic radius, of (bio)polymers, nanoparticles or even small molecules. In this work, we describe an extension to current data analysis schemes that allows size polydispersity to be quantified for an arbitrary sample, thereby significantly enhancing the potentiality of Taylor dispersion analysis. The method is based on a cumulant development similar to that used for the analysis of dynamic light scattering data. Specific challenges posed by the cumulant analysis of Taylor dispersion data are discussed, and practical ways to address them are proposed. We successfully test this new method by analyzing both simulated and experimental data for solutions of moderately polydisperse polymers and polymer mixtures.

**Introduction**

Taylor dispersion analysis (TDA) is an absolute, straightforward and fast method for determining diffusion coefficients $D$, and hydrodynamic radii $R_h$. The principle of TDA is based on the dispersion of a narrow solute band in an open tube under Poiseuille laminar flow.[1-2] Due to the parabolic velocity profile, the solutes move with different velocities depending on their position in the tube cross section. The Taylor dispersion is due to the combination of the dispersive velocity profile with molecular diffusion that redistributes the molecules over the cross section of the tube. The most common way to perform TDA relies on recording the solute's concentration profile as a function of time at a given spatial position. The determination of the diffusion coefficient is then based on the experimental determination of the temporal variance of the elution profile.[3-10] TDA was first applied on long open tubes in gaseous phase[3], then in liquids.[4-6] More recently, capillary electrophoresis instruments that allow the solute concentration profile to be recorded at a given location in a narrow capillary (diameter ~50 µm) were shown to be particularly well suited for TDA.[11] TDA is applicable to solutes of virtually any size from angstroms to sub-micron and of any nature (small molecules, macromolecules, dendrimers, nanoparticles, liposomes…).[12-16] TDA can also be implemented in non-aqueous phase for the characterization of hydrophobic compounds.[17] Since it is an absolute method, neither calibration nor knowledge of the sample concentration are required. Only a few nL of sample is usually injected (~1% of the total capillary volume), which makes this method suitable for the analysis of biological and pharmaceutical compounds such as proteins[18-20] or drug delivery systems[21], where sample availability is an important issue.

In the case of mixtures of solutes, or polydisperse samples, the average diffusion coefficients that can be measured by TDA based on the determination of the variance of the sample peak depends on the nature of the detector (mass concentration- or molar concentration sensitive detector).[22] In a previous work, we demonstrated that, for the commonly used mass concentration sensitive detector, TDA leads to a harmonic weight-averaged diffusion coefficient, and therefore to the weight-averaged hydrodynamic radius.[22]

Beyond the average value of the diffusion coefficient obtained by integration of the taylorgram, no general method exists to quantify size dispersion for polydisperse samples. Some attempts were proposed for specific cases such as bimodal mixtures.[23] The theory dealing with three-component systems has been proposed by Price[24], however it is somehow complex as it involves diffusion cross-terms. Boyle et al.[25] determined diffusion coefficients of pauci- and polydisperse poly(styrene sulfonate) samples by studying the variation of the peak width with the carrier velocity, by flow injection analysis. However, since they operated in conditions where the injected product is eluted from the tube in a time smaller than the characteristic time of diffusion across the tube section, their approach does not rely on Taylor's analysis of dispersion and does not correspond to TDA. Mes et al.[7] reported a comparison of different methods, including TDA, for the determination of diffusion coefficients of polydisperse synthetic copolymers. Fitting the taylorgram by a sum of Gaussian functions should, in principle, allow the determination of the distribution of the diffusion coefficients of the mixture. This approach has been applied to synthetic mixtures of up to six-mers.[26] However, it should be emphasized that this method requires the knowledge of the exact number of components, and is hardly applicable for too large a number of components, since the least-square fit of the taylorgram profile becomes numerically ill-conditioned.

In this work, we present a new data analysis scheme for (moderately) polydisperse samples, termed the cumulant method[27]. This method relies on the analogy with the cumulant analysis widely used in dynamic light scattering (DLS)[28], another popular size-characterization method. Briefly, the cumulant analysis consists in a change of variables against which the raw data (here the temporal taylorgram) are plotted, leading to a linear behavior of the data in the case of a monodisperse sample. For polydisperse samples, deviations from linearity are observed, which are quantified via a second-order polynomial fit of the curved data. As we will show it in the following, the coefficients of the linear and quadratic terms of the polynomial fit are related to well-defined moments of the distribution of the diffusion coefficients, $D$. This allows for a quantitative determination of both the average size and the polydispersity of the sample. Furthermore, the average size obtained from the cumulant analysis corresponds to a different moment of the distribution of $D$, as compared to the harmonic average usually measured by TDA. Based on this observation, we introduce a new quantitative indicator of polydispersity that is particularly robust with respect to data noise.

The paper is organized as follows. In the next section, the theoretical bases of the cumulant analysis are presented. The cumulant method is then applied to simulated taylorgrams generated from size exclusion chromatography (SEC) distributions, to demonstrate the validity and interest of this approach. In this section, we also introduce the new polydispersity index based on the ratio between the usual harmonic average and that issued from the cumulant analysis. Finally, the cumulant method is applied to experimental taylorgrams obtained for polystyrenesulfonate standards of various molecular weights and their mixtures.

## THEORY

### Transformation from molar mass distribution to diffusion coefficient distribution for polydisperse polymer samples

The mass-weighted probability distribution function (PDF) of the molar mass $M$ of a given polymer sample is defined as:

$$P_M(M) = \frac{M\rho_M(M)}{\int_0^\infty M\rho_M(M)dM} \tag{1}$$

where $\rho_M(M)dM$ is the molar concentration of the species with molar mass between $M$ and $M+dM$. It is convenient to convert $P_M(M)$ to the mass-weighted PDF of diffusion coefficients, because Taylor dispersion data are more naturally expressed as a function of diffusion coefficients. The mass-weighted PDF of $D$ is defined by:

$$P_D(D) = \frac{M(D)\rho_D(D)}{\int_0^\infty M(D)\rho_D(D)dD} \tag{2}$$

with $M(D)$ the molar mass of the species with diffusion coefficient $D$ and $\rho_D(D)$ their molar concentration distribution function. The molar mass and the diffusion coefficient are related by

$$D = \frac{k_B T}{\eta}\left(\frac{5N_A}{324\pi^2 K}\right)^{1/3} M^{-\frac{1+a}{3}} \tag{3}$$

where $k_B$ is the Boltzmann's constant, $T$ the absolute temperature, $\eta$ the solvent viscosity, $N_A$ Avogadro's number, and where $K$ and $a$ are the Mark-Houwink parameters relating the intrinsic viscosity to $M$ through $[\eta] = KM^a$. The PDF of $D$ is then obtained using eq 3 and the standard PDF transformation law:

$$P_D(D) = \left[P_M(M)\left|\frac{dM}{dD}\right|\right]_{M=M(D)} \tag{4}$$

**Taylor dispersion analysis: theoretical bases.**

For a sample solution of a single component of molar concentration $\rho$, the time evolution of the signal $S$ measured in a Taylor dispersion experiment is Gaussian:

$$S(t) = CM\rho\sqrt{D} \exp\left[-\frac{(t-t_0)^2 12D}{R_c^2 t_0}\right] \tag{5}$$

where $t_0$ is the peak time, $R_c$ the radius of the capillary, $C$ an instrumental constant. The molar mass $M$ has been introduced in eq. 5 since usually the detector response is proportional to the mass concentration. The extension to the case of a molar-concentration sensitive detector may be obtained by replacing the molar mass $M$ by unity in eq.5 and in eqs. 6 and 11 below. Note that eq. 5 only holds if $\frac{Dt_0}{R_c^2} \geq 1.4$ to ensure that the average detection time is larger than the characteristic diffusion time of the solutes in the capillary cross section.[1, 12] Equation 5 is also only valid if the axial diffusion can be neglected compared to the Taylor dispersion contribution. This latter condition is fulfilled if $\frac{R_c u}{D} \geq 69$, where $u$ is the linear velocity of the mobile phase.[1, 12] Furthermore, corrections due to the finite injection time have been neglected since the injected volume for TDA experiments was always lower than ~1% of the capillary volume.[10] The generalization of eq 5 to a polydisperse sample reads:

$$S(t) = \int_0^\infty CM(D)\rho_D(D)\sqrt{D} \exp\left[-\frac{(t-t_0)^2 12D}{R_c^2 t_0}\right] dD \tag{6}$$

assuming that the sample is diluted enough for the cross-diffusion between sample components to be negligible.[26] Using eq 2, the mass-weighted average diffusion coefficient for a polydisperse sample is:

$$\overline{D} = \int_0^\infty D P_D(D) dD \tag{7}$$

where here and in the following we denote by an overbar mass-weighted averages obtained from the distribution functions. Note that $\overline{D}$ is not directly accessible in a Taylor dispersion experiment. However, an average diffusion coefficient may be easily obtained from the temporal variance of $S(t)$, by defining:

$$\langle D \rangle_T \equiv \frac{R_c^2 t_0}{24} \frac{\int_0^\infty S(t) dt}{\int_0^\infty S(t)(t-t_0)^2 dt} \tag{8}$$

We shall denote $\langle D \rangle_T$ as the "Taylor average" of the diffusion coefficient. Experimentally, the upper limit of the integrals in eq 8 is replaced by the largest available time (provided that $S(t)$ has decayed to $\approx 0$ at the end of the experiment), or by the peak time $t_0$. By replacing $S(t)$ in eq 8 by the r.h.s. of eq 6 and using eq 7, one recognizes that the Taylor average is the (mass-weighted) harmonic mean of $D$:

$$\langle D \rangle_T = \left[\overline{D^{-1}}\right]^{-1} \leq \overline{D} \tag{9}$$

The last inequality of eq 9 follows from the general properties of the harmonic mean and shows that the Taylor average weighs more the species with a small $D$ (i.e. a large $M$ or $R_h$), as compared to the arithmetic mean. The inequality reduces to equality for a monodisperse sample. Note that when considering the hydrodynamic radius (which is inversely proportional to the diffusion coefficient), Taylor dispersion analysis leads to the arithmetic weight-average hydrodynamic radius, to which all species contribute proportionally to their relative weight.[22]

**Theory of cumulant method**

We introduce here the cumulant method for analyzing Taylor dispersion data. As stated in introduction, the aim of the cumulant approach is to linearize the data for a monodisperse sample. In the case of a polydisperse sample, the deviation to linearity will give a measure of the polydispersity. Note that in a previous work[29] a similar linearized representation was used, but no attempts were made to quantify polydispersity by analyzing the deviations from a linear behavior. It is convenient to normalize $S(t)$ by its peak value, by defining

$$s(t) \equiv S(t)/S(t_0) = \int_0^\infty f(D)\exp\left[-\frac{(t-t_0)^2 12D}{R_c^2 t_0}\right]dD \tag{10}$$

with

$$f(D) = M(D)\rho_D(D)\sqrt{D} \Big/ \int_0^\infty M(D)\rho_D(D)\sqrt{D}\,dD \tag{11}$$

For future use, we define the "Gamma average" of an arbitrary quantity $A$ depending on $D$ as

$$\langle A \rangle_\Gamma = \int_0^\infty f(D)A(D)dD. \tag{12}$$

In the spirit of a cumulant expansion, we now show that for a moderately polydisperse sample $s(t)$ may be written as the taylorgram for monodisperse objects with diffusion coefficient $\langle D \rangle_\Gamma$, the Gamma average of $D$, times a correction term that depends on the width of the size distribution. The diffusion coefficient of a particular species in a polydisperse sample may be written as

$$D = \langle D \rangle_\Gamma + \delta D \tag{13}$$

where by definition $\langle \delta D \rangle_\Gamma = 0$. By inserting eq 13 in eq 10, one obtains:

$$s(t) = \exp\left[-\frac{(t-t_0)^2 12\langle D \rangle_\Gamma}{R_c^2 t_0}\right]\int_0^\infty f(D)\exp\left[-\frac{(t-t_0)^2 12\delta D}{R_c^2 t_0}\right]dD, \tag{14}$$

where terms independent of $D$ have been factored in front of the integral. We simplify eq. 14 by assuming $\frac{(t-t_0)^2 12\delta D}{R_c^2 t_0} \ll 1$, corresponding to moderate polydispersity and/or $t$ close to the peak time ($t-t_0 \to 0$). Under this assumption, the exponential in the integral of eq 14 may be replaced to a good approximation by its Taylor expansion up to the second order. Using the normalization of *f(D)*, eq 11, and the definition of $\delta D$, eq 13, one finds that terms of order $\delta D$ cancel out, leading to:

$$s(t) = \exp\left[-\frac{(t-t_0)^2 12 \langle D \rangle_\Gamma}{R_c^2 t_0}\right]\left[1 + \frac{1}{2}\left(\frac{(t-t_0)^2 12}{R_c^2 t_0}\right)^2 \langle \delta D^2 \rangle_\Gamma + ...\right] \quad (15)$$

We take the natural logarithm of eq 15 and, in the same spirit of the approximation applied to eq. 14, we further use $\ln(1+x) \approx x$ for $x = \frac{1}{2}\left(\frac{(t-t_0)^2 12}{R_c^2 t_0}\right)^2 \langle \delta D^2 \rangle_\Gamma \to 0$, finally obtaining:

$$\ln[s(t)] = -\Gamma_1 (t-t_0)^2 + \frac{\Gamma_2}{2}(t-t_0)^4 + ... \quad (16)$$

with

$$\Gamma_1 = \frac{12}{R_c^2 t_0}\langle D \rangle_\Gamma \quad (17)$$

$$\Gamma_2 = \left(\frac{12}{R_c^2 t_0}\right)^2 \langle \delta D^2 \rangle_\Gamma. \quad (18)$$

Equation 16 is the central theoretical result of this paper: it shows that the logarithm of the taylorgram may be expanded in a cumulant series in $(t-t_0)^2$. The first cumulant, $\Gamma_1$, is directly related to the Gamma average of the diffusion coefficient given by eq 17. Note that $\langle D \rangle_\Gamma$ differs from both the weight-averaged arithmetic mean and the Taylor average of *D*. Using eqs 7, 9 and 12, one has:

$$\langle D \rangle_\Gamma = \frac{\overline{D^{3/2}}}{\overline{D^{1/2}}} \geq \overline{D} \geq \langle D \rangle_T. \quad (19)$$

The equalities hold only for monodisperse samples; for polydisperse samples the Gamma average is biased towards the species with large *D* (*i.e.* small $R_h$ or *M*), as compared to both the arithmetic and the Taylor averages. The second cumulant, $\Gamma_2$, is related to the width of the PDF of the diffusion coefficient. A convenient non-dimensional parameter that quantifies the relative width of the distribution –and thus the sample polydispersity- is the Gamma-averaged relative variance, defined by:

$$\frac{\Gamma_2}{\Gamma_1^2} = \frac{\langle D^2 \rangle_\Gamma - \langle D \rangle_\Gamma^2}{\langle D \rangle_\Gamma^2} = \frac{\dfrac{\overline{D^{5/2}}}{\overline{D^{1/2}}} - \left(\dfrac{\overline{D^{3/2}}}{\overline{D^{1/2}}}\right)^2}{\left(\dfrac{\overline{D^{3/2}}}{\overline{D^{1/2}}}\right)^2} \qquad (20)$$

## EXPERIMENTAL SECTION
**Chemical and polymers**

Borax (disodium tetraborate decahydrate) was purchased from Prolabo (Paris, France). The water used to prepare all buffers was further purified with a Milli-Q-system from Millipore (Molsheim, France). The borate buffers were directly prepared by dissolving the appropriate amount of borax in water. Standards of poly(styrene sulfonate) (PSS, weight average molar masses $M_w$ 1.29×10³, 5.19×10³, 29×10³, 145×10³, 333×10³ g/mol) were purchased from Polymer Standards Service (Mainz, Germany). The polydispersity index of the PSS is below 1.2. The degree of sulfonation of the PSS is higher than 90%. All PSS standards were provided with the PDF of $M$ (numerical data, derived from SEC data, were obtained from Polymer Standards Services on simple request). The Mark-Houwink parameters ($K$, $a$) of polystyrenesulfonate are determined in 80 mM sodium borate buffer at 25°C as explained in Supporting Information (See Figure SI-1).

**Taylor dispersion analysis**

Taylor dispersion analysis (TDA) experiments were performed on a PACE MDQ Beckman Coulter (Fullerton, CA) apparatus. Capillaries were prepared from bare silica tubing purchased from Composite Metal Services (Worcester, United Kingdom). Capillary dimensions were 40 cm (30 cm to the detector) × 50 µm I.D. New capillaries were conditioned with the following flushes: 1 M NaOH for 30 min, 0.1 M NaOH for 30 min and water for 10 min. Before sample injection, the capillary was filled with the buffer (80 mM borate buffer, pH 9.2, 8.9 $10^{-4}$ Pa.s viscosity). PSS samples were dissolved in the buffer at 0.5 g/L. Between two TDA analyses, the capillary was successively flushed with: (i) water (50 psi, 1 min); (ii) 1M NaOH (50 psi, 2 min) and (iii) buffer (50 psi, 3 min). Solutes were monitored by UV absorbance at a wavelength of 200 nm. Sample injection was performed hydrodynamically on the inlet side of the capillary (0.3 psi, 9s; ~1% of the capillary volume). Mobilization pressures of 2 psi were applied with buffer vials at both ends of the capillary. Pressure ramp time was 15 s. The elution time was systematically corrected for the delay in the application of the pressure by substracting 7.5 s (half-time of the pressure ramp) to the observed (recorded) elution time.[22] The temperature of the capillary cartridge was set at 25 °C

## RESULTS AND DISCUSSION

**Cumulant analysis of simulated taylorgrams**

To illustrate the cumulant method, we analyze simulated Taylor dispersion data generated for both moderately polydisperse polymers and a mixture of two polymer batches with different molar mass. The simulated *s*(*t*) is obtained using eqs 10 and 11, where $M(D)\rho_D(D)$ is calculated from the SEC PDF of *M* via eqs 4 and 2. We use realistic conditions (sampling time 0.25 s, $t_0 \approx 77$ s) and we add to the numerical data a random noise drawn from a Gaussian distribution, with a standard deviation equal to 0.0001, the noise level being determined by comparison with the typical noise level seen in experimental data. The quantities obtained from the data analysis are systematically compared to their theoretical values directly computed using the input $P_D(D)$.

More specifically, the expected value of $\langle D \rangle_T$, $\Gamma_1$, $\langle D \rangle_\Gamma$, $\Gamma_2$ and $\Gamma_2/\Gamma_1^2$ are computed from the PDF of $D$ using eqs 9, 17, 19, 18, and 20, respectively. Table SI-T1 gathers some of these parameters. Note that the Mark-Houvink coefficients required in eqs 3 and 4 were obtained as shown in Figure SI-1.

Figure 1a displays $P_D(D)$, obtained from the SEC data, for the PSS 5190 sample. This PDF was used to generate the taylorgram shown in Figure 1b. Figure 1c shows the cumulant representation of the same data. For the cumulant analysis, we only take into account absorbance data collected during the raising slope, since experiments (presented in the next section) show that this is typically cleaner, most likely because some polymers eventually stick to the capillary walls and pollute the falling slope of the taylorgram. Note that in the cumulant representation, Figure 1c, the data deviate slightly from a straight line, thus revealing that the sample is not strictly monodisperse. Such deviations would be difficult to be appreciated in the traditional representation of Figure 1b.

When analyzing a taylorgram with the cumulant method, two practical issues need to be addressed: first, only data close to the maximum of $s(t)$ should be considered, so that higher-order terms that were truncated in eq 16 are actually negligible. However, it is clear that reducing too much the range of the cumulant fit would lead to large errors, due to data noise. How should then the optimum cutoff level be determined ? Second, the peak time $t_0$ needs to be known with good precision, since any error in its determination would spuriously modify the values of $\Gamma_1$ and $\Gamma_2$ issued from a cumulant fit, as we shall show in the following. Note that while the first point is also typical of DLS data analysis, the latter is specific to Taylor dispersion data. To address both issues, we perform a series of cumulant fits by varying systematically the cutoff level and by testing various guess values of $t_0$ in a small interval around the experimental peak time (determined *e.g.* as the maximum of $s(t)$ or through a parabolic fit around such extremum). We then inspect the cutoff-dependence of both cumulants for the chosen guess values of $t_0$ and determine accordingly the best cutoff level and peak time. To perform realistic tests of this procedure, the software used to generate the numerical taylorgrams adds a small random number to the user-input value of $t_0$, so that the actual peak time is not known at the time of data analysis. The actual peak time is stored by the software, so that the effectiveness of the procedure can be verified *a posteriori*.

Typical results of this analysis are presented in Figure 2, for the data shown in Figure 1. The top panel shows $\Gamma_1$ as a function of the cutoff level, for various guess values of the peak time, as indicated by the labels. If the guess value is too small, the data sharply increase as the cutoff level is raised. The opposite trend occurs when the guess value is too large. Note that large deviations are observed even when the guess values depart from the true peak value by just a fraction of a second. This demonstrates the importance of determining with good accuracy the peak value using this procedure. The best choice, $t_0 = 76.804$ s, is determined such that the first cumulant has only a very mild dependence on the cutoff level. We recall that the actual value was not known before analyzing the data. The optimum value found here is very close to the actual value used to generate the data ($t_0 = 76.85$ s), thus validating the proposed fitting procedure. Once the optimum $t_0$ is fixed, the first cumulant is finally determined by extrapolating a linear fit to the $\Gamma_1$ *vs* cutoff data at the highest cutoff value, *i.e.* cutoff = 1. Only data for a cutoff value $\geq 0.5$ are considered in the fit, and data at very high cutoff levels are excluded if significant deviations from the general trend are observed, due to data noise. For the data shown here and once the optimum peak time has been fixed, $\Gamma_1$ depends only very weakly on the cutoff level and the contribution of data noise is negligible. Accordingly, the choice of the best cutoff level is not crucial. We find however that this may be important in real data, especially for the second cumulant and the relative variance, as we shall show it in the experimental tests section. Figures 2b and 2c show the same kind of analysis for $\Gamma_2$ and for the relative variance. A linear fitting procedure similar to that used for $\Gamma_1$ is applied to the relative variance (solid line in Figure 2c). Once the relative variance is determined, its value and that of $\Gamma_1$ are used to determine $\Gamma_2$, as shown by the arrow in Figure 2b. For both the relative variance and $\Gamma_2$ the general trend as a function of the cutoff level and the choice of the peak time is similar to that for the first cumulant, Figure 2a. Note however that here the dependence on $t_0$ is even more marked. Indeed, a bad choice of the peak time may even lead to a *negative* second cumulant which, in view of eq 18, is unphysical. Table SI-T1 given in supporting information compares the theoretical values of $\Gamma_1$ and $\Gamma_2$ calculated by integration (eqs 17 and 12 for $\Gamma_1$, and eqs 18 and 12 for $\Gamma_2$) of the PDF of $D$ derived from the SEC distributions to the values obtained by the cumulant approach (eq 16) applied to the simulated taylorgrams. For $\Gamma_1$, the theoretical and simulated values are in excellent agreement (average relative difference of 1.4% for all the simulated monomodal samples) demonstrating the practical feasibility and usefulness of the cumulant analysis. Relative differences are, on average, slightly higher for bimodal polymer mixtures (up to 10%). Similar comparisons were also performed on simulated data obtained with a noise level ten time larger (*i.e.* a Gaussian random noise with standard deviation 0.001). Results on $\Gamma_1$ were not affected by these higher levels of noise.

Concerning the second cumulant $\Gamma_2$, its relative error (with respect to the theoretical value issued from the SEC distributions) is typically 8-9 times larger than that on first cumulant. Decreasing the noise level down to 0.0001 does not reduce significantly this error. Therefore,

Figures 2b and 2c and Table SI-T1 in the Supporting Information highlight how delicate it may be to extract precise information on the size distribution from $\Gamma_2$.

An alternative way of quantifying polydispersity may be obtained by comparing the Gamma average $<D>_\Gamma$, issued from $\Gamma_1$, to the Taylor average, $<D>_T$. Indeed, eq. 19 shows that the Gamma average weighs more the larger species, as compared to the Taylor average. While for a strictly monodisperse sample the two averages coincide, for polydisperse samples $\langle D \rangle_\Gamma > \langle D \rangle_T$, the difference being larger for a greater polydispersity. Motivated by this observation, we introduce a polydispersity index *PI* based on the ratio between $<D>_\Gamma$ and $<D>_T$. Several definitions are *a priori* possible; we choose in particular a definition based on the notion of an "equivalent log-normal" PDF. We introduce a log-normal PDF as:

$$P_D(D) = P_{LN}(D) \equiv \frac{1}{\gamma D \sqrt{2\pi}} \exp\left[-\frac{(\ln D - \beta)^2}{2\gamma^2}\right] \tag{21}$$

where the parameters $\beta$ and $\gamma$ are chosen in such a way that $\langle D \rangle_T$ and $\langle D \rangle_\Gamma$ calculated using the log-normal PDF, eq 21, coincide with those obtained for the real sample. Using eqs 8, 19, and 21, one finds:

$$\beta = \frac{1}{3}\ln\langle D \rangle_\Gamma + \frac{2}{3}\ln\langle D \rangle_T \tag{22}$$

$$\gamma = \sqrt{\frac{2}{3}\ln\frac{\langle D \rangle_\Gamma}{\langle D \rangle_T}} \tag{23}$$

A natural definition of the polydispersity index PI is then

$$PI \equiv \gamma^2 = \frac{2}{3}\ln\frac{\langle D \rangle_\Gamma}{\langle D \rangle_T} \ . \tag{24}$$

The choice of a log-normal distribution presents several advantages: a log-normal PDF is characterized by just two parameters, $\beta$ and $\gamma$; for such a distribution, $\langle D \rangle_T$ and $\langle D \rangle_\Gamma$ can be calculated analytically, yielding eqs 22 and 23; a log-normal distribution is often a good approximation for the distribution of (monomodal) real samples such as polymers or colloidal particles; a log-normal PDF is invariant under a power-law change of variable, so that if any of the diffusion coefficient $D$, the hydrodynamic radius $R_h$ or the molecular weight $M$ are distributed log-normally, the same will also apply to the two other quantities. Of course, for an arbitrary shape of $P_D$ the equivalent log-normal distribution will in general differ from the true PDF of $D$; nonetheless, the equivalent log-normal PDF allows one to get a sense of the actual distribution of the diffusion coefficients simply using the Taylor and Gamma averages issued from a straightforward data analysis.

Figure 3 compares the actual PDF used to generate the data (dotted line, same data as in Figure 1a) to the equivalent log-normal distribution obtained using eqs 22 and 23 (blue solid line): for this monomodal sample, $P_{LN}(D)$ captures very well both the position of the peak of $P_D$ and its width. We note for completeness that the parameters $\beta$ and $\gamma$ may also be deduced from the first two cumulants. Using eqs 17, 18 and 21, one finds

$$\beta = \ln \Gamma_1 - \ln\left(1 + \frac{\Gamma_2}{\Gamma_1^2}\right) + \ln \frac{R_c^2 t_0}{12} \tag{25}$$

$$\gamma = \sqrt{PI} = \sqrt{\ln\left(1 + \frac{\Gamma_2}{\Gamma_1^2}\right)} \tag{26}$$

Figure 3 shows the equivalent log-normal distribution obtained from the first two cumulants (red dashed line). In view of the sensitivity of $\Gamma_2$ to data noise discussed above, however, we generally prefer to determine the equivalent log-normal distribution using the Taylor and Gamma averages, rather than $\Gamma_1$ and $\Gamma_2$.

In Figure SI-3 of the Supporting Information we show that the principle of the cumulant analysis also applies to the case of a bimodal distribution of diffusion coefficients, a situation often encountered in experiments, *e.g.* in the monitoring of polymerization processes.[23] Results for various bimodal samples are summarized in Table SI-T1 of the Supporting Information, where we show that the value of $\Gamma_1$ expected from the SEC distributions is indeed recovered to within 10%. The equivalent log-normal distributions issued from the Taylor and Gamma averages of *D* are compared to the PDF of *D* used to generate the data in the Supporting Information (Figure SI-3c). While by construction the equivalent log-normal PDF cannot capture the bimodal nature of the input distribution, we emphasize that $P_{LN}$ still provides useful indications on the range of *D* covered by the actual distribution.

**Cumulant analysis of experimental taylorgrams**

To fully test the cumulant method proposed here, we have analyzed experimental Taylor dispersion data obtained for solutions of PSS of various molecular weights, and for a mixture of two polymers (as those tested for the simulated data).

Figure 4a shows the experimental taylorgrams for diluted solutions of PSS5190 and PSS29k, as well as for an equimass mixture of the two polymers. As expected, the signal from the smallest polymer exhibits the fastest decay, while that for the mixture lays in between those of the monomodal samples. Figure 4b shows a cumulant plot of the same signals, obtained from the raising slope of *s*(*t*). The lines are second order cumulant fits to the data in the range that was actually used for the cumulant analysis. As already observed for the simulated data (see Figure SI-3), the short time behavior of ln[*s*] is dictated by both polymer species, while at larger $(t - t_0)^2$ the taylorgram for the mixture follows the behavior expected for the largest species, as demonstrated by the fact that the slope of the triangles and the circles is essentially identical at large values of $(t - t_0)^2$.

As for the simulated samples, we determine the optimal choices of the peak time and the cutoff level by examining plots of $\Gamma_1, \Gamma_2$ and the ratio $\Gamma_2/\Gamma_1^2$ as a function of both $t_0$ and the cutoff, as exemplified in the Supporting Information for the solution of PSS5190 (Figure SI-5). As a general trend, we find that the determination of $\Gamma_1$ is as robust as for the simulated data. By contrast, the determination of the second cumulant is less straightforward, since $\Gamma_2$ varies significantly and non-monotonically as the cutoff level tends to one. As a consequence, it is also difficult to determine unambiguously the ratio $\Gamma_2/\Gamma_1^2$, which is an indicator of the sample polydispersity via eq 20. These difficulties stem from the relatively narrow size distribution of the monomodal sample, combined with the unavoidable experimental noise. As a consequence, the curvature of $\ln[s]$ vs $(t-t_0)^2$ is very modest and the noise significantly affects its determination. Under these circumstances, an estimate of the sample polydispersity via the polydispersity index introduced in eq 24 and the equivalent log-normal distribution is particularly valuable, since the PI is obtained from the two averages $\langle D \rangle_T$ and $\langle D \rangle_\Gamma$, which are more robust than $\Gamma_2$ with respect to data noise.

Figure 5 shows the equivalent log-normal distributions for the two monomodal samples PSS5190 and PSS29k, and for the mixture. For the monomodal samples, the equivalent-log normal distributions are in good agreement with $P_D$ as estimated from the SEC distributions (dotted lines), although they appear to be somehow broader. This difference may stem from the different analytical method used in SEC as compared to Taylor dispersion. However, a similar trend was also observed for the simulated data, see Figure 3. It is therefore likely that this discrepancy is due to the shape of the actual PDF, which differs from a log-normal distribution. Figure 5c shows the PDFs for the bimodal solution. As already observed for the simulated data (see Figure SI-3), the log-normal PDF does not capture (by construction) the bidisperse nature of the sample, but nonetheless it provides a good estimate of the overall width of the distribution of diffusion coefficients and of its position. Interestingly, we find that for the mixture the equivalent log-normal distributions calculated using the Gamma and Taylor averages or the two cumulants essentially coincide, in contrast to what was observed for the monomodal samples, Figures SI 7a-c. This trend is also observed for the simulated samples (compare Figure 3 to Figure SI-3c). It is most likely due to the fact that for the relatively broad distributions associated with bimodal samples the curvature of $\ln[s]$ *vs* $(t-t_0)^2$ is more pronounced than for the monomodal solutions, thereby allowing one to measure the second cumulant in a more reliable way.

The same approach was used to analyze the experimental taylorgrams obtained with all the PSS studied in this work having a weight average molar mass of 1.29k, 5.19k, 29k, 145k, 333k and some of their (50/50 w/w) bimodal mixtures. The resulting $P_D$ distributions of PSS monomodal samples are displayed in Figure SI-6. As expected, the distributions obtained from $\Gamma_1, \Gamma_2$ (red dashed line) appear to be in general in a less good agreement with the SEC distribution (dotted black line) as compared to what was obtained for the simulated data (see Figure SI-2). If one considers the $\langle D \rangle_T$, $\Gamma_1$ approach, the agreement with the SEC distribution is much better, as we observed in the case of simulated taylorgrams. The relative difference on $\Gamma_1$ between the experiments and the value expected from the SEC distribution appears to be reasonable, with an average relative difference of ~30% as seen in Table SI-T1. It should be emphasized that this difference may stem not only from uncertainties in the analysis of the experimental taylorgrams, but also from similar errors in the determination of the actual size distribution by SEC. The log normal distributions represented in SI clearly shows that even when the difference between the results obtained for $\Gamma_1$, from the taylorgrams and the SEC is as high as 50%-60% (corresponding to the largest observed differences, see PSS 1.29k and 145k in Table SI-T1), the log-normal distributions obtained from experimental taylorgrams are still very informative. In particular, the range of $D$ over which the $P_D$ distributions significantly depart from zero is captured reasonably well by the equivalent log-normal distribution. This trend is also confirmed by the bimodal distributions displayed in Figure SI-7 that may be compared to the results for the simulated taylorgrams in Figure SI-4.

As a final remark, we note that the applicability of the cumulant analysis relies on fulfilling the condition $\frac{(t-t_0)^2 12\delta D}{R_c^2 t_0} \ll 1$ (see eqs. 14 and 15), which in turns sets an upper limit $\delta D \ll \frac{R_c^2 t_0}{12(t-t_0)^2}$ on the maximum polydispersity that can be reliably measured. In the case of our experiments where the taylorgram is sampled at a rate of 4 Hz, we take $(t-t_0) \geq 2$ s in order to perform the fit on at least 8 data points, thus finding $\delta D \approx 10^{-5}$ cm$^2$ s$^{-1}$, five times larger than the largest $<D>_T$.

**Conclusion**

In this work, it has been demonstrated that the cumulant approach, which is commonly used in dynamic light scattering for polydispersity analysis, can be similarly used for Taylor dispersion analysis. This approach was applied to the analysis of moderately polydisperse polymer samples and bimodal mixtures of these samples. The cumulant analysis of taylorgrams requires first the accurate determination of the average elution time $t_0$ by a systematic graphical representation that can be numerically automated. It has been shown that the polydispersity of the sample is more precisely obtained by using the first order cumulant value $\Gamma_1$ combined with $\langle D \rangle_T$ instead of using the two cumulant order parameters $\Gamma_1$ and $\Gamma_2$. This can be explained by a better determination and better precision on $\langle D \rangle_T$ that is obtained by direct integration of the taylorgram, as compared to the second order $\Gamma_2$ parameter. $\Gamma_1$ and $\langle D \rangle_T$ give an estimation of the polydispersity of the sample. It has then been proposed to use the log normal distribution having the same polydispersity as the real sample distribution as a way to quantify polydispersity. These log normal distributions were favorably compared to distributions obtained by SEC. Knowing the simplicity of implementation of TDA, and its wide applicability, we believe that this approach could be routinely used for the characterization of samples of virtually any size from angstrom to sub-micron, and of any nature.

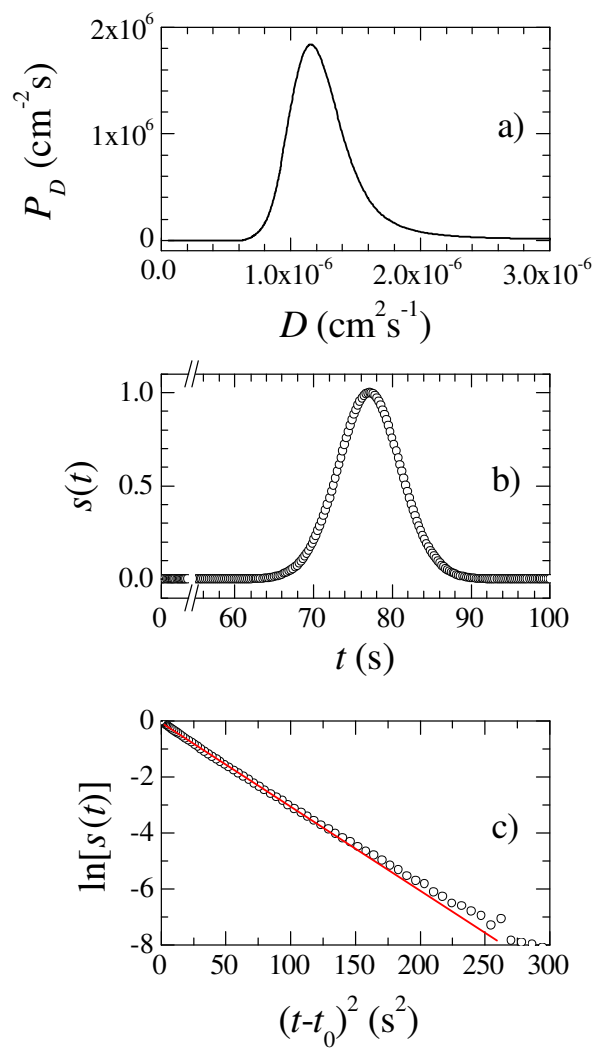

**Figure 1.** a) Probability distribution function of the diffusion coefficient $D$ used as an input in eq 6 to generate the taylorgram shown in b). c) Symbols: cumulant plot of the raising slope of the taylorgram shown in b). The red line is a linear fit to the data in the range $s(t) > 0.22$. The modest deviations with respect to a linear behavior indicate that the sample is only slightly polydisperse.

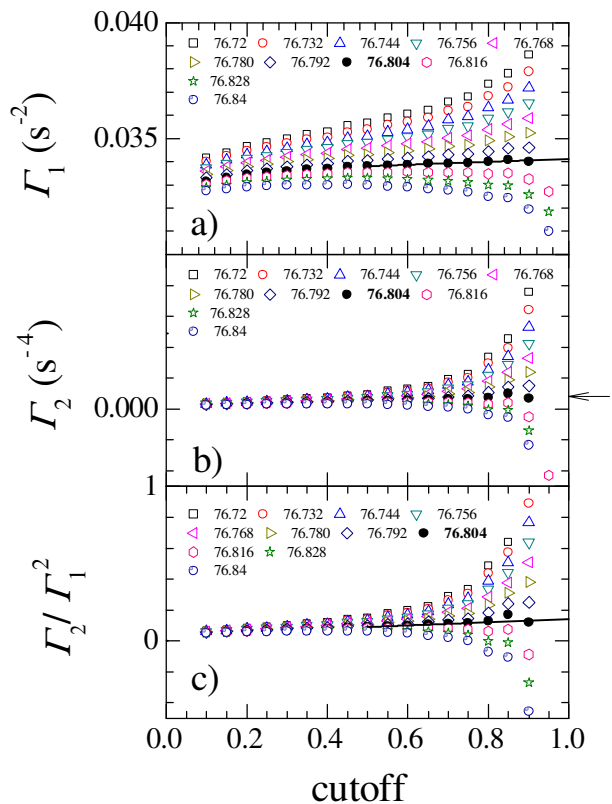

**Figure 2.** First cumulant (a), second cumulant (b), and relative variance (c) as a function of the cutoff level in fitting the data shown in Figure 1c. The curves are labeled by the guess value (in s) of the peak time used in the cumulant fit, eq 16. The filled symbols and the bold label refer to the chosen peak time, $t_0$ = 76.804 s. The lines in a) and c) are linear fits to the data with the optimum $t_0$: their intersection with the line cutoff = 1 yields the desired values of $\Gamma_1$ and $\Gamma_2/\Gamma_1^2$. The arrow in Figure 2b shows the value retained for $\Gamma_2$, based on the results of 2a and 2c.

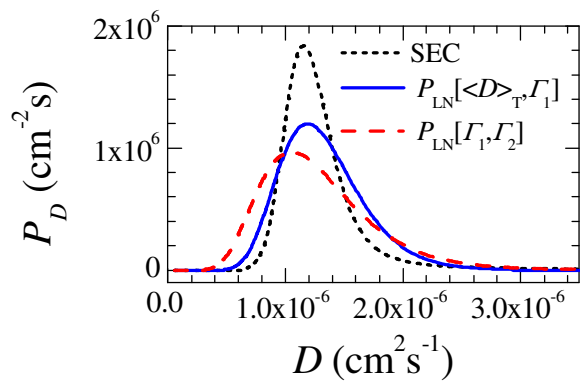

**Figure 3**. Probability distribution function of the mass-averaged diffusion coefficient for a mildly polydisperse polymer, PSS 5190. Dotted line: PDF obtained from the SEC data and used to simulate the taylorgram data. Blue solid line (resp., red dashed line): equivalent log-normal distribution issued from the Taylor and Gamma averages (resp., from the first two cumulants) determined by analyzing the simulated taylorgram.

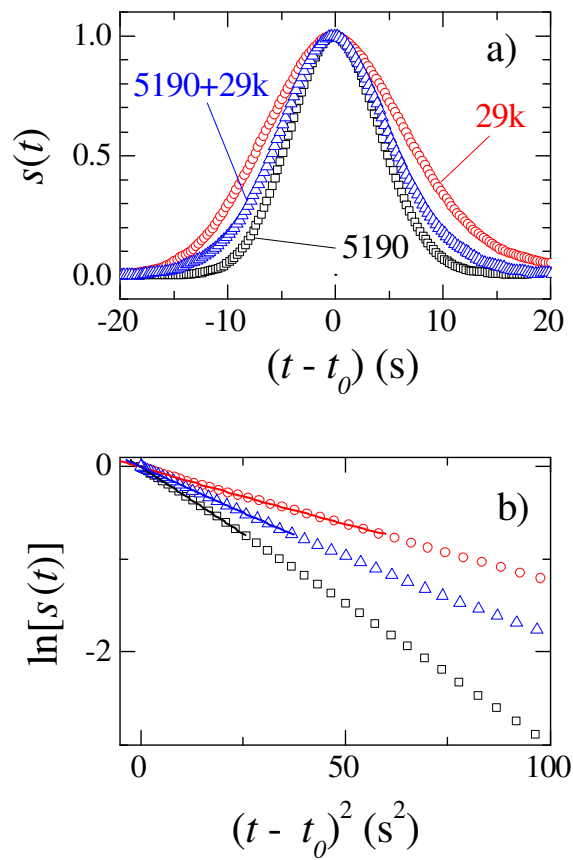

**Figure 4**. Experimental taylorgrams (a) measured for moderately polydisperse PSS 5190k (squares) and 29k (circles), and for an equimass mixture of both polymers (triangles). Cumulant plots (b) obtained from the raising slope of the data shown in a). The lines are second-order cumulant fits to the data, in the optimum range of $t$ determined through the analysis illustrated by Figure SI-5.

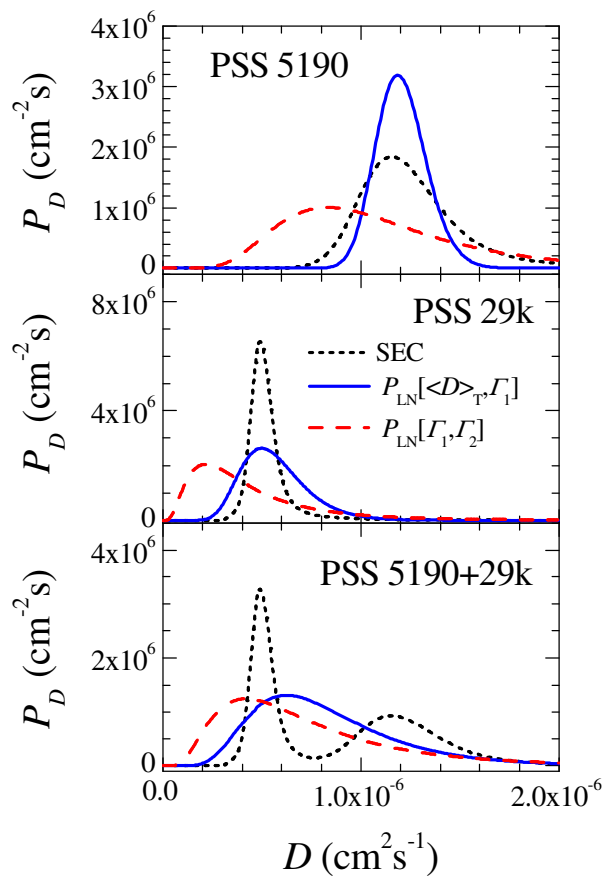

**Figure 5**. Mass-weighted probability distribution functions of the diffusion coefficient for two monomodal samples (PSS5190, (a), and PSS 29k, (b)) and an equimass mixture of both polymers (c). In all panels, the dotted curves are $P_D$ as estimated by SEC, while the blue solid (resp., red dashed) lines are the equivalent log-normal distributions retrieved from the analysis of the experimental taylorgrams, using the Taylor and Gamma averages (resp., the first two cumulants).

AUTHOR INFORMATION

**Corresponding Authors**

*luca.cipelletti@univ-montp2.fr

*hcottet@univ-montp2.fr

**Author Contributions**

The manuscript was written through contributions of all authors. All authors have given approval to the final version of the manuscript.

**ACKNOWLEDGMENT**

H.C. gratefully acknowledges the support from the Région Languedoc-Roussillon for the fellowship "Chercheurs d'Avenir" and from the Institut Universitaire de France (2011-2016).

**SUPPORTING INFORMATON AVAILABLE**

Supporting information includes:

- Figure SI-1: Molar mass dependence of the Taylor-averaged diffusion coefficient, $<D>_T$, for the monomodal PSS polymers investigated;

- Figure SI-2: Equivalent log-normal distributions and SEC distributions of $D$ for the five monomodal samples, data issued from the analysis of the simulated taylorgrams;

- Figure SI-4: Equivalent log-normal distributions and SEC distributions of $D$ for the three bimodal samples, data issued from the analysis of the simulated taylorgrams;

- Figure SI-6: Equivalent log-normal distributions and SEC distributions of $D$ for the five monomodal samples, data issued from the analysis of the experimental taylorgrams;

- Figure SI-7: Equivalent log-normal distributions and SEC distributions of $D$ for the three bimodal samples, data issued from the analysis of the experimental taylorgrams;

- Table SI-T1: values of $<D>_T$, $\Gamma_1$ and $\Gamma_2$ as calculated from the SEC distributions, and as obtained from the analysis of the simulated and experimental taylorgrams.

This information is available free of charge via the Internet at http://pubs.acs.org/

**For TOC only**

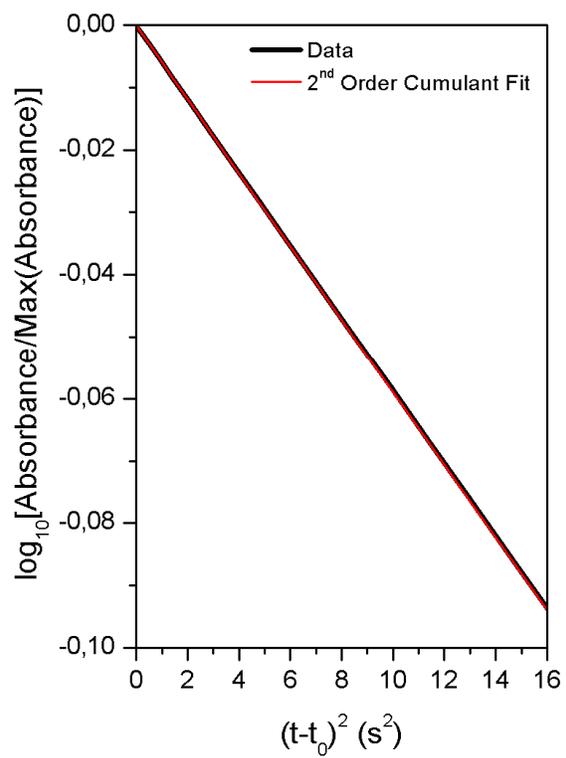
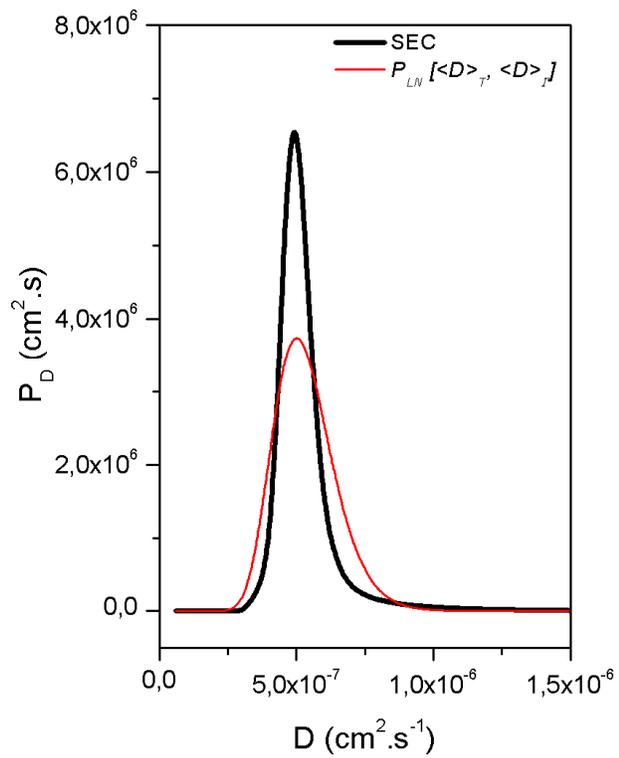

*Supporting Information to*

# Polydispersity analysis of Taylor dispersion data: the cumulant method

*Luca Cipelletti[1], Jean-Philippe Biron[2], Michel Martin[3], Hervé Cottet[2]*

[1]Laboratoire Charles Coulomb (L2C, UMR 5221 CNRS-Université Montpellier 2), Place Eugène Bataillon, F-34095 Montpellier Cedex 5, France

[2]Institut des Biomolécules Max Mousseron (IBMM, UMR 5247 CNRS-Université de Montpellier 1- Université de Montpellier 2), Place Eugène Bataillon, CC 1706, F-34095 Montpellier Cedex 5, France

[3]Ecole Supérieure de Physique et de Chimie Industrielles, Laboratoire de Physique et Mécanique des Milieux Hétérogènes (PMMH, UMR 7636 CNRS, ESPCI-ParisTech, Université Pierre et Marie Curie, Université Paris-Diderot), 10 rue Vauquelin, F-75231 Paris Cedex 05, France

*TABLE OF CONTENTS*

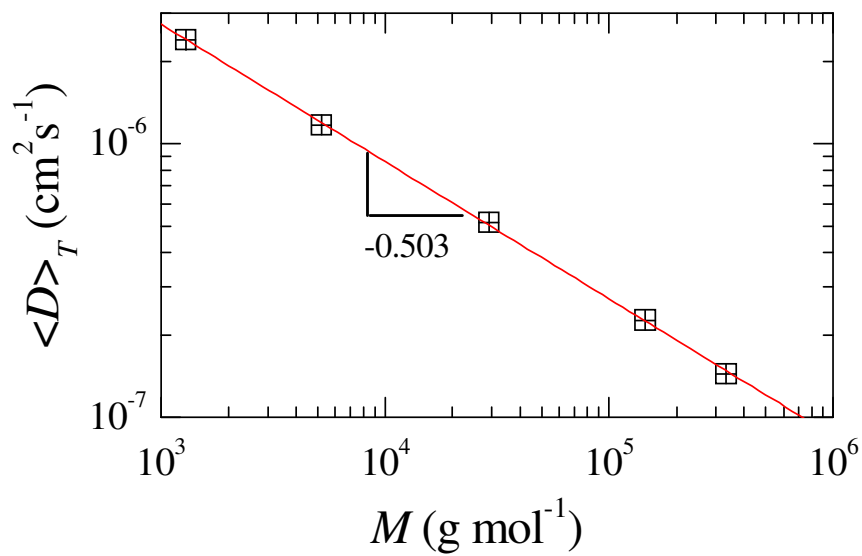

**Figure SI-1**: Experimental Taylor-averaged diffusion coefficient, $<D>_T$, as a function of the nominal molar mass, for the five PSS polymers used in this study. The data are very well fitted by the power law $\langle D \rangle_T = cM^d$, with $c = 8.83 \times 10^{-5}$ (c.g.s. units) and $d = -0.503$. The Mark - Howink parameters $a$ and $K$ are obtained from $c$ and $d$ by comparing the power law shown here to eq. 3, yielding $a = 0.5083$, $K = 0.11$ (c.g.s. units). The values thus obtained are used to convert the SEC mass distribution to that of the diffusion coefficient, as explained in the theoretical section (see eqs 3 and 4).

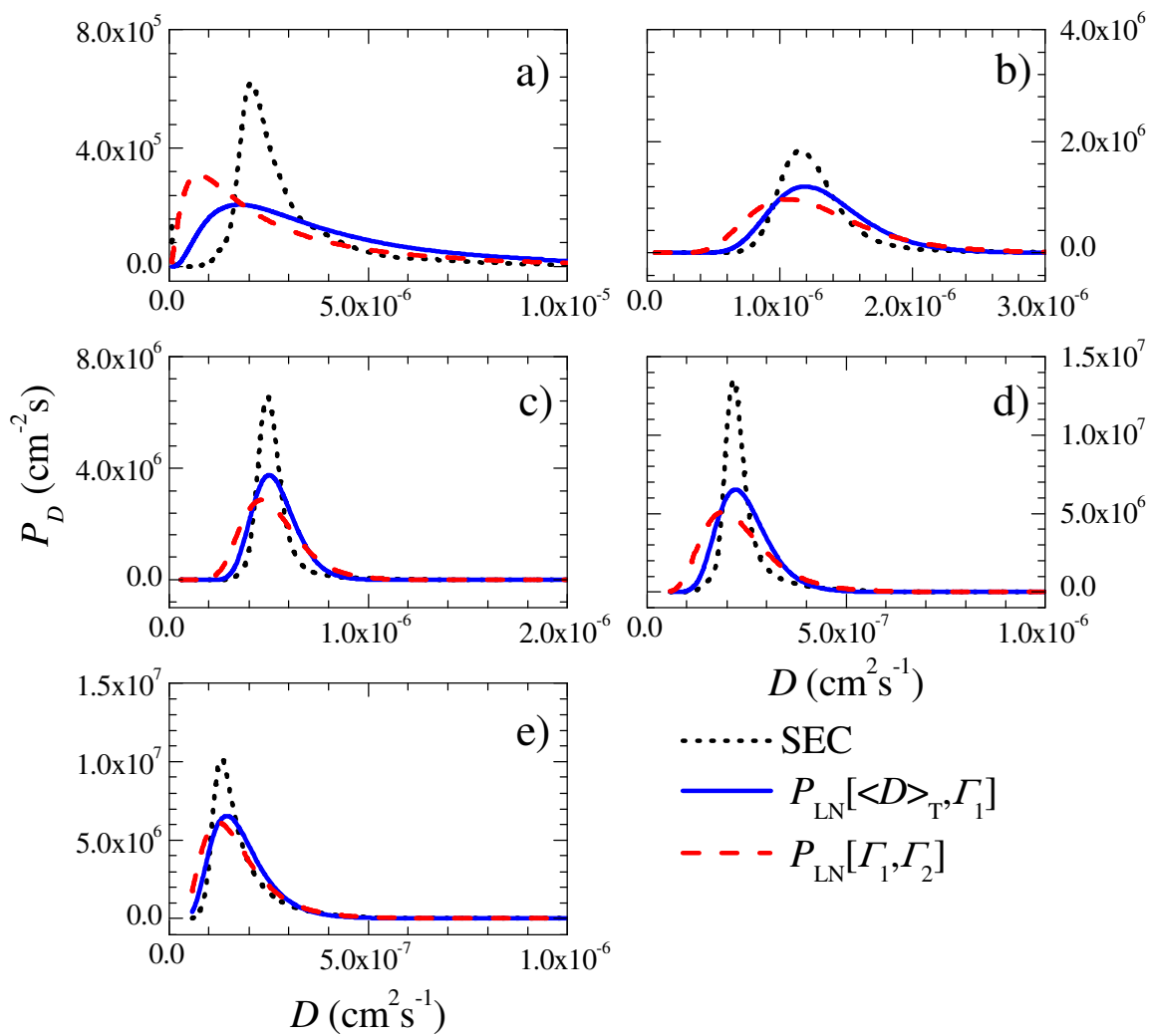

**Figure SI-2:** Probability distribution function of the diffusion coefficient $D$ calculated from the SEC data (dotted black line) and equivalent log-normal distributions obtained from the simulated taylorgrams, using the Taylor-averaged diffusion coefficient and $\Gamma_1$ (blue solid line), or $\Gamma_1$ and $\Gamma_2$ (red dashed line). For all simulated taylorgrams shown here, a random noise with a standard deviation of 0.0001 was added to the signal. Monomodal samples: a) PSS 1290, b) PSS 5190 (same as in Figure 3 of the main text), c) PSS 29k, d) PSS 145k, e) PSS 333k.

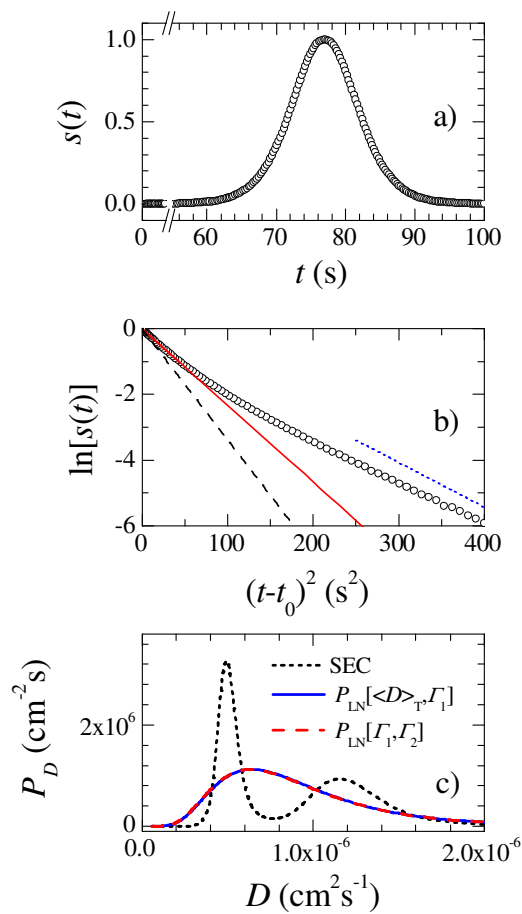

**Figure SI-3:** Simulated taylorgram for an equimass mixture of PSS 5190 and PSS 29k (a). Cumulant plot of the same data (b). Note that the short-time behavior of ln[$s$] is determined by both species, and not only by the contribution associated with the smaller polymer, whose taylorgram decays faster. This is shown by the difference between the solid line (first order cumulant fit) and the black dashed line (behavior expected if only the smallest species, PSS 5190, was present). At large $t$, the slope of ln[$s(t)$] is close to that expected for the sample PSS 29k alone (blue dotted line). (c) Mass-weighted PDF of the diffusion coefficient $D$ used as an input to generate the taylorgram shown in (a) and (b) (dotted line), together with the equivalent log-normal distributions obtained from the Taylor and Gamma averages (blue solid line) and the first two cumulants (red dashed line), respectively.

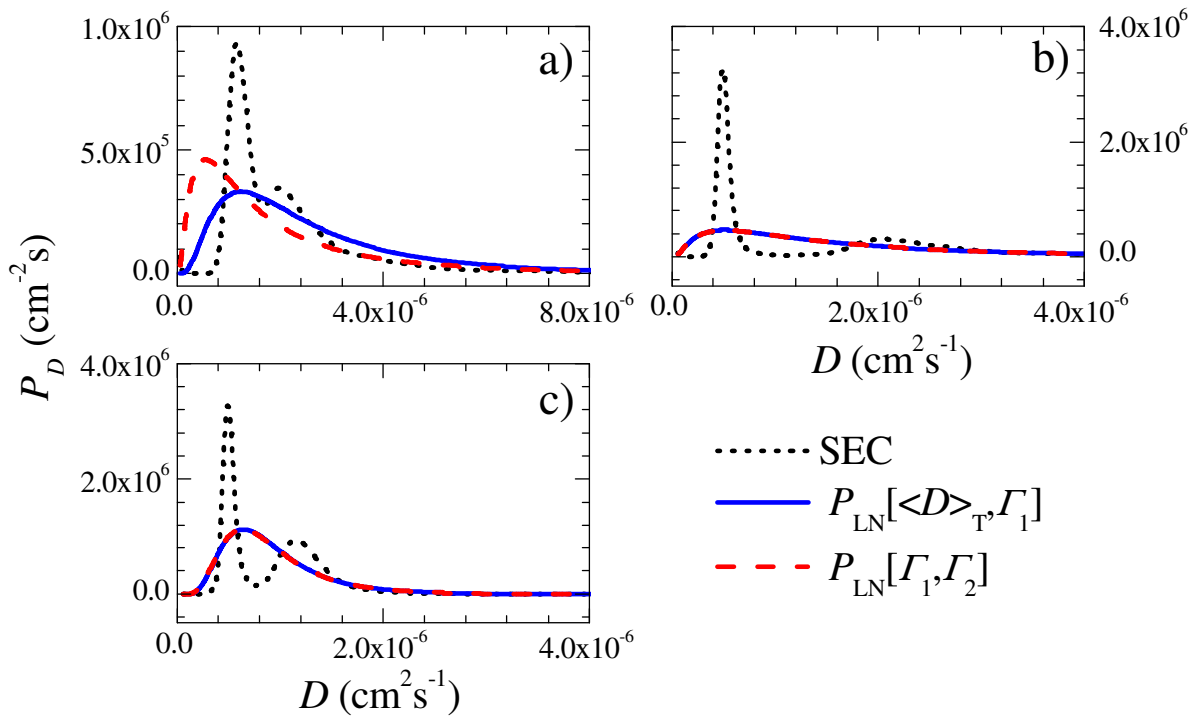

**Figure SI-4:** Probability distribution function of the diffusion coefficient $D$ calculated from the SEC data (dotted black line) and equivalent log-normal distributions obtained from the simulated taylorgrams, using the Taylor-averaged diffusion coefficient and $\Gamma_1$ (blue solid line), or $\Gamma_1$ and $\Gamma_2$ (red dashed line). For all simulated taylorgrams shown here, a random noise with a standard deviation of 0.0001 was added to the signal. Equimass mixtures of : a) PSS 1290 and PSS 5190, b) PSS 1290 and PSS 29k (same as Figure SI-3c), c) PSS 5190 and PSS 29k.

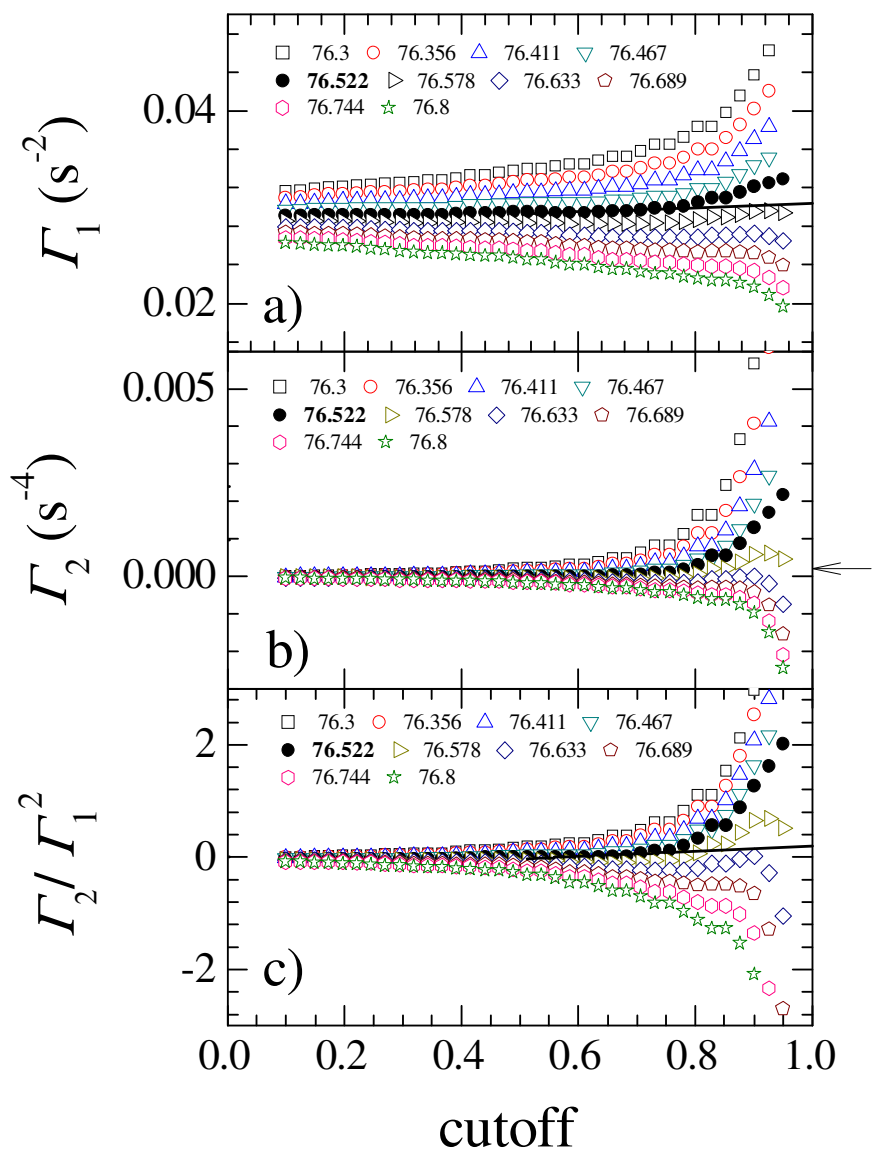

**Figure SI-5**. First cumulant (a), second cumulant (b), and relative variance (c) as a function of the cutoff level in fitting the experimental PSS 5190 data shown in Figure 4b. The curves are labeled by the guess value (in s) of the peak time used in the fit. The filled symbols and the bold label refer to the chosen peak time. The lines in a) and c) are linear fits to the data with the optimum $t_0$: their intersection with the the line cutoff = 1 yields the desired values of $\Gamma_1$ and $\Gamma_2/\Gamma_1^2$. The arrow in b) shows the value retained for $\Gamma_2$, based on the results of a) and c).

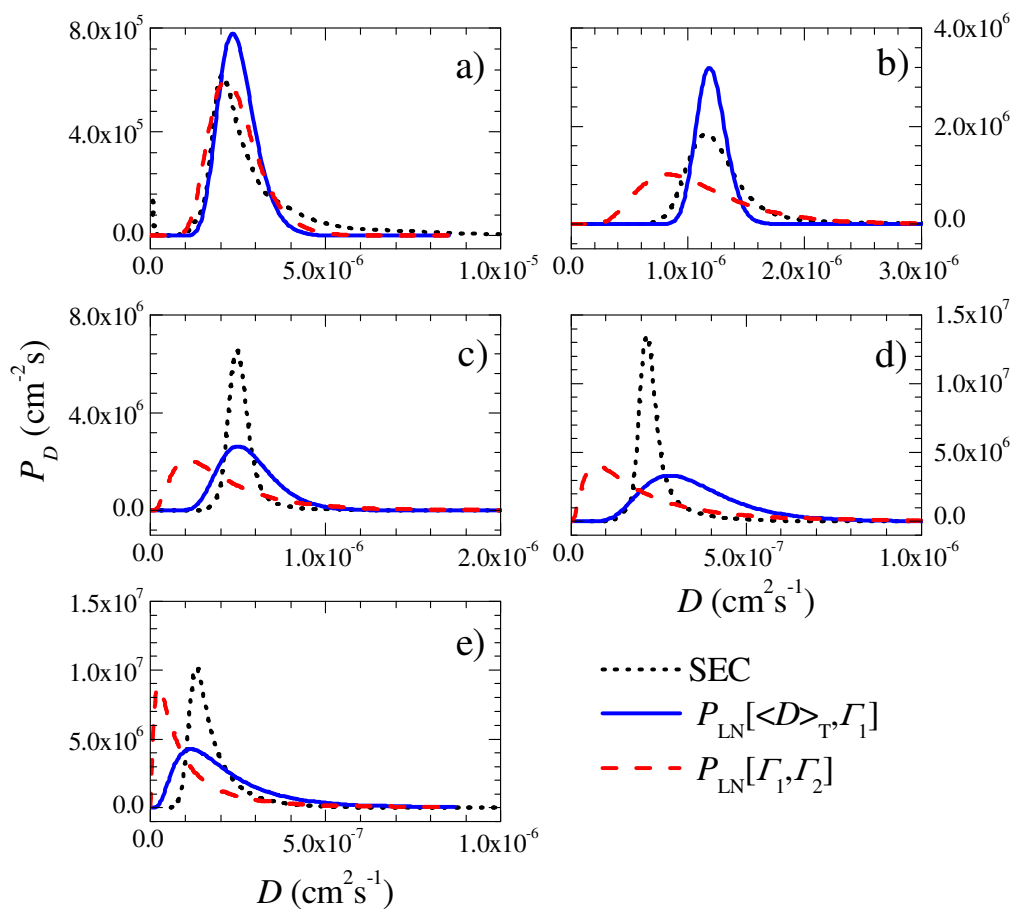

**Figure SI-6:** Probability distribution function of the diffusion coefficient $D$ calculated from the SEC data (dotted black line) and equivalent log-normal distributions obtained from the experimental taylorgrams, using the Taylor-averaged diffusion coefficient and $\Gamma_1$ (blue solid line), or $\Gamma_1$ and $\Gamma_2$ (red dashed line). Monomodal samples: a) PSS 1290, b) PSS 5190 (same as Figure 5a of the main text), c) PSS 29k (same as Figure 5b of the main text), d) PSS 145k, e) PSS 333k.

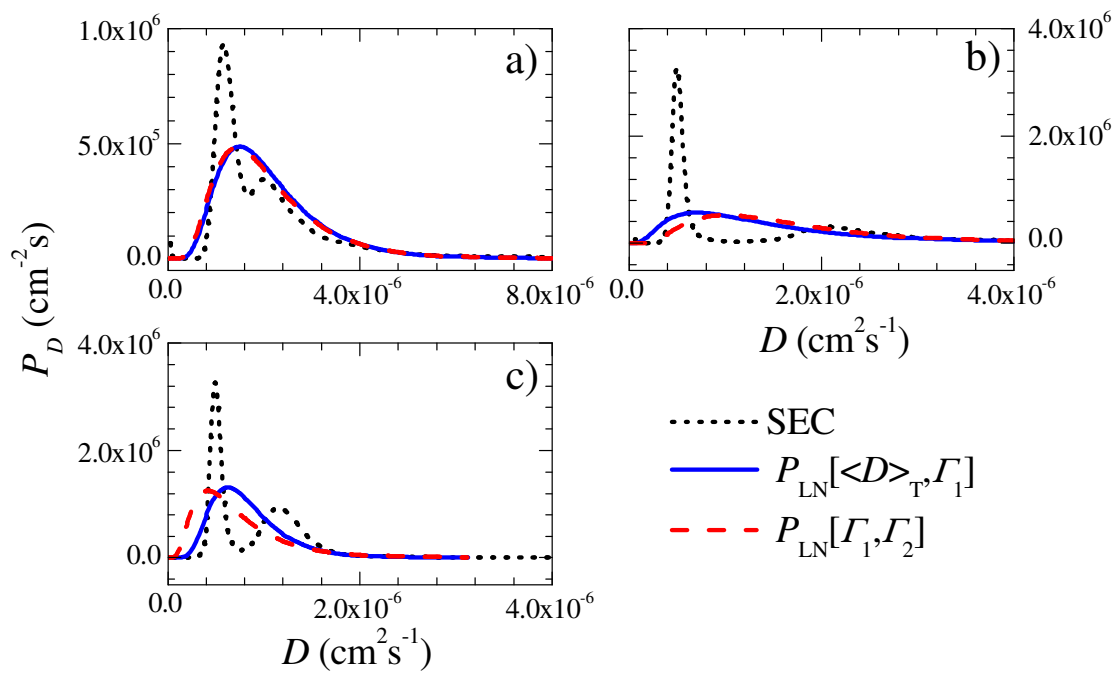

**Figure SI-7:** Probability distribution function of the diffusion coefficient $D$ calculated from the SEC data (dotted black line) and equivalent log-normal distributions obtained from the experimental taylorgrams, using the Taylor-averaged diffusion coefficient and $\Gamma_1$ (blue solid line), or $\Gamma_1$ and $\Gamma_2$ (red dashed line). Equimass mixtures of : a) PSS 1290 and PSS 5190, b) PSS 1290 and PSS 29k, c) PSS 5190 and PSS 29k (same as Figure 5c of the main text).

|  |  | PSS1290 | PSS5190 | PSS29000 | PSS145000 | PSS333000 | PSS1290 + PSS5190 | PSS1290 + PSS290000 | PSS5190 + PSS29000 |
|---|---|---|---|---|---|---|---|---|---|
| SEC | $<D>_T$ (cm$^2$ s$^{-1}$) | 2.34E-06 | 1.22E-06 | 5.11E-07 | 2.29E-07 | 1.55E-07 | 1.61E-06 | 8.39E-07 | 7.21E-07 |
|  | $\Gamma_1$ (s$^{-2}$) | 1.47E-01 | 3.39E-02 | 1.36E-02 | 2.35E-03 | 4.87E-03 | 1.03E-01 | 1.09E-01 | 2.60E-02 |
|  | $\Gamma_2$ (s$^{-4}$) | 5.84E-02 | 1.44E-04 | 1.44E-05 | 5.07E-07 | 5.93E-06 | 3.90E-02 | 4.53E-02 | 1.91E-04 |
| Simulated | $<D>_T$ (cm$^2$ s$^{-1}$) | 2.40E-06 | 1.23E-06 | 5.12E-07 | 2.30E-07 | 1.56E-07 | 1.63E-06 | 8.45E-07 | 7.24E-07 |
|  | $\Gamma_1$ (s$^{-2}$) | 1.55E-01 | 3.41E-02 | 1.37E-02 | 2.37E-03 | 4.90E-03 | 9.26E-02 | 9.70E-02 | 2.60E-02 |
|  | $\Gamma_2$ (s$^{-4}$) | 4.56E-02 | 1.66E-04 | 1.63E-05 | 7.22E-07 | 6.29E-06 | 1.37E-02 | 1.65E-02 | 1.89E-04 |
| Experimental | $<D>_T$ (cm$^2$ s$^{-1}$) | 2.40E-06 | 1.19E-06 | 5.21E-07 | 2.29E-07 | 1.43E-07 | 1.69E-06 | 9.30E-07 | 6.89E-07 |
|  | $\Gamma_1$ (s$^{-2}$) | 6.41E-02 | 3.04E-02 | 1.47E-02 | 3.83E-03 | 6.89E-03 | 5.88E-02 | 5.50E-02 | 2.30E-02 |
|  | $\Gamma_2$ (s$^{-4}$) | 3.74E-04 | 1.92E-04 | 1.44E-04 | 1.89E-05 | 1.22E-04 | 1.03E-03 | 1.48E-03 | 2.59E-04 |

**Table SI-T1**: Values of $<D>_T$, $\Gamma_1$ and $\Gamma_2$ as calculated from the SEC distributions, and as obtained from the cumulant analysis of the simulated and experimental taylorgrams. Theoretical $\Gamma_1$ and $\Gamma_2$ values (three SEC lines) were calculated by integration (eqs. 17 and 12 for $\Gamma_1$, and eqs. 18 and 12 for $\Gamma_2$) of the PDF of $D$ obtained from SEC distributions. Simulated values were obtained by the cumulant approach (eq. 16) on the simulated taylorgrams constructed from the SEC distributions. Experimental values were obtained by the cumulant approach (eq. 16) on the experimental taylorgrams. Bimodal mixtures were prepared on the basis of a 50/50 w/w mixture.